# Equilibrium Probability Distribution for Number of Bound Receptor-Ligand Complexes

*Tuhin Chakrabortty[*], Manoj M Varma[*#]*
*[*]Center for Nano Science and Engineering, Indian Institute of Science, Bangalore, India*
*[#] Robert Bosch Centre for Cyber-Physical Systems, Indian Institute of Science, Bangalore, India*

**Abstract**
The phenomenon of molecular binding, where two molecules, referred to as a receptor and a ligand, bind together to form a ligand-receptor complex, is ubiquitous in biology and essential for the accurate functioning of all life-sustaining processes. The probability of a single receptor forming a complex with any one of $L$ surrounding ligand molecules at thermal equilibrium can be derived from a partition function obtained from the Gibbs-Boltzmann distribution. We extend this approach to a system consisting of $R$ receptors and $L$ ligands to derive the probability density function $p(r; R, L)$ to find $r$ bound receptor-ligand complexes at thermal equilibrium. This extension allows us to illustrate two aspects of this problem which are not apparent in the single receptor problem, namely, a) a symmetry to be expected in the equilibrium distribution of the number of bound complexes under exchange of $R$ and $L$ and b) the number of bound complexes obtained from chemical kinetic equations has an exact correspondence to the maximum probable value of $r$ from the expression for $p(r; R, L)$. We derive the number fluctuations of $r$ and present a practically relevant molecular sensing application which benefits from the knowledge of $p(r; R, L)$.

## 1. Introduction

The phenomenon of molecular binding, where two molecules, generally referred to as a receptor and a ligand, bind together to form a ligand-receptor complex, occurs ubiquitously in biological systems and is essential for the accurate functioning of many life-sustaining processes ranging from DNA transcription to immune response [1, 2]. The ligand-receptor complex formation is a reversible process, which means that while a ligand molecule can associate with a receptor molecule to form a ligand-receptor complex, the converse process, namely the dissociation or breaking down of a ligand-receptor complex into one each of the ligand and receptor molecules can also happen simultaneously. In general, the rates of these two competing processes would be different leading to one process dominating over the other. As we will see in the next section of this article, thermodynamic equilibrium is reached when the rates of these two processes become equal. The equilibrium state is also a steady state where the number of the bound ligand-receptor complexes remains constant over time due to the two equal and opposite processes cancelling each other. While living organisms are non-equilibrium systems which exchange energy and matter with the surroundings, as illustrated beautifully by Rob Philips, the separation of time-scales allows a large number of biological processes involving molecular binding to be viewed as processes in equilibrium [3]. Due to this ubiquitous and critical role of molecular binding in biological functions, we would like to be able to answer questions such as, how many ligand-receptor complexes do we expect to find at equilibrium after we start with $R$ receptors and $L$ ligands. As we will see in the next paragraph, there are a number of different ways to answer this question and related questions such as the magnitude of fluctuations in the number of ligand-receptor complexes at equilibrium.

One of the ways to answer the question on the number of ligand-receptor complexes to be expected at equilibrium, starting from $R$ receptors and $L$ ligands can be answered by considering the molecular binding as a reversible chemical reaction and writing down the rate equations by invoking the law of mass action[4]. This approach is worked out in section two of this article. This approach provides not only the number of ligand-receptor complexes at equilibrium but also how the number of complexes change as a function of time from the initial value of zero. A question then arises, namely, if we perform $N$ independent experiments each with $R$ receptors and $L$ ligands and measure the number of ligand-receptor complexes at equilibrium for each of these experiments, will we always get the same number as predicted by the reaction rate approach? As we might expect, the answer is no. There will be variation in the number of ligand-receptor complexes seen in the different experiments which can be quantitatively characterized by the variance or standard deviation of the measured values. The variation in the measured number of complexes arises not just because of experimental uncertainties which may be present but also due to a more fundamental reason, namely, that the binding of a receptor and a ligand molecule to form the complex as well as its dissociation at a later point in time are stochastic events which follow certain probability distributions. As a result, even if the experimental uncertainties were reduced to zero hypothetically, one would still find a distribution of values for the number of ligand-receptor complexes even if one performed identical experiments repeatedly. Therefore, an estimate for the magnitude of the variation is essential to obtain a fundamental understanding of noise in biological systems [5, 6, 7, 8] or in a more applied context to understand and optimize performance of biosensors. Biosensors are devices which rely on molecular binding to detect or quantify the presence of specific molecules in human samples such as blood. The amount of these specific molecules, referred to as biomarkers, indicate the presence of diseases such as cancer [9]. One may like to know the fluctuations in the biosensor signal caused due to the stochastic nature of molecular binding in diagnostic applications, which are generally performed under equilibrium conditions. The magnitude of these fluctuations set a fundamental limit on the lowest While the reaction rate approach can tell us the number of ligand-receptor complexes at equilibrium, it cannot give us any information regarding the statistical distribution of this quantity and related parameters of interest such as the magnitude of fluctuations (quantified by parameters such as the variance or the standard deviation) around the mean number of complexes. It is in this context that the application of statistical physics to this problem helps us answer these questions in detail.

We know that equilibrium statistical mechanics allows us to calculate the distribution of particles into the various states of a given system. These states may be discrete or continuous and are characterized by a specific energy level of the state. By enumerating all the states of a given system appropriately scaled by their thermodynamic weights, one can obtain the partition function of the system which then yields the probability distribution of particles in these states. In the molecular binding problem, one could consider the unbound ligand and receptor molecules as one state and the bound complex as another state forming a two-state system and calculate the partition function. This approach has been exactly followed to derive the equilibrium occupation probability of a single receptor surrounded by $L$ ligand molecules [3, 6, 10, 11]. The equilibrium occupation probability of the receptor is the probability that the receptor will be bound by one of the $L$ ligand molecules in equilibrium. The resultant formula is shown to be equivalent to the one which is obtained from the chemical reaction rate equations [12]. This problem serves as a good illustration of how statistical mechanics links macroscopic parameters such as the equilibrium dissociation constant to the microscopic, molecular level details such as the energy level difference between bound and unbound states.

Although the problem with one receptor and $L$ ligands serves to illustrate the application of equilibrium statistical mechanics to solve the binding problem, in many real situations we have more than one receptors. For instance, a cell may have multiple receptors on its surface and biosensor certainly has millions of receptors which are used to bind a ligand target. Therefore, we would like to extend the solution of the one receptor problem into one which has $R$ receptors. The formal statement of the problem is to calculate the probability distribution of the number of receptor-ligand complexes $r$, under thermal equilibrium conditions, given $R$ receptors and $L$ ligands. If the binding of ligands to receptors can be considered as independent events, then the answer is simply $R$ times the probability of binding a single receptor by $L$ ligands which has been solved previously. The assumption of independence is a little tricky because the occupation of one receptor by a ligand implies that for the next complex formation there is one less receptor and one expects that the probabilities of binding would change. However, are there situations where independence of events can be justified? To answer such questions systematically, one way forward is to write the partition function for the system and obtain the required probability distribution. An alternate method of deriving the probability distribution of receptor-ligand complex is from the master equation describing the binding process which is Markovian, and accounting for transitions into and out of the state with $r$ receptor-ligand complexes [13]. One can obtain the partition function from a recurrence relation extracted from the master equation. The advantage of this method is that in addition to the equilibrium probability distribution of the number of complexes, one can also obtain the temporal evolution of this distribution. Here we illustrate the first method of solution.

In this article we derive the partition function by direct enumeration and from it the probability density function $p(r; R, L)$, which is the probability to find $r$ bound receptor-ligand complexes at thermal equilibrium with the initial receptor and ligand concentrations of $R$ and $L$ respectively. The article is divided into the following sections. In section 2, we derive the number of bound complexes $r_{eq}$, at equilibrium using the conventional approach based on balancing of forward and backward reaction rates. Here we show that the expression for the number of bound receptors from the binding kinetic equation $(b_{eq})$ is symmetric with respect to the exchange of $R$ and $L$ and discuss why this symmetry is to be expected in this problem. In section 3, we present the partition function for a system consisting of $R$ receptors and $L$ ligands which can bind to form complexes. We show that this expression and the probability density function $p(r; R, L)$ which follows from it is also symmetric (invariant) under exchange of $R$ and $L$. In section 3, we provide explicit expressions for $p(r; R, L)$ and show that the expression for $r_{eq}$ derived from the reaction rate approach corresponds to $r_m$ which maximizes $p(r; R, L)$, and not to the expectation value $\langle r \rangle$. This indicates that $p(r; R, L)$ is a skewed distribution, although the skew is not very significant. In section 4, we show that under certain regimes $p(r; R, L)$, which is a rather complicated expression, can be approximated by a binomial distribution. We then derive the number fluctuations of the bound complex and finally in section 5, we connect this problem to a practical situation involving the sensing of biomolecules in the context of the extremely important application of early diagnosis of disease conditions.

**2. The reaction rate approach**

Let us consider a system with $R$ receptors and $L$ ligands. The ligands can bind to the receptors reversibly with binding rate $k_{on}$ and unbind with a rate $k_{off}$. The general reaction kinetic equation specifying the rate of formation of complexes $b$ follows from the law of mass action as,

$$\frac{db}{dt} = k_{on}(R - b)(L - b) - k_{off}b \tag{1}$$

Here, the first term in the RHS represents the instantaneous rate of free ligand and free receptor molecules associating to form a complex while the second term represents the instantaneous rate of dissociation of existing ligand-receptor complexes. At equilibrium, these two opposing reaction rates are equal and the RHS becomes equal to zero from which we see that the number of complexes $b_{eq}$ at equilibrium satisfies the quadratic equation below,

$$k_{on}(R - b_{eq})(L - b_{eq}) = k_{off}b_{eq} \tag{2}$$

Solving Eq. (2) for $b_{eq}$, we get

$$b_{eq} = \frac{(R + L + K_D) \pm \sqrt{(R + L + K_D)^2 - 4RL}}{2}, \tag{3}$$

where $K_D = \frac{k_{off}}{k_{on}}$.

Equation (3) describes the general solution for the number of bound complexes at steady state indicating two possible values of $b_{eq}$ at equilibrium. However, we know that $b_{eq}$ cannot exceed the maximum of $L$ or $R$. Therefore, the only physically acceptable solution of $b_{eq}$ is,

$$b_{eq} = \frac{(R + L + K_D) - \sqrt{(R + L + K_D)^2 - 4RL}}{2} \tag{4}$$

For special cases e.g. $R \ll L$ and $R \gg L$, one can modify Eq (2) by ignoring the depletion in the number of receptor and ligand molecules as shown in Eq. 5(a) and Eq. 5(b) respectively,

$$k_{on}(R - b_{eq})L = k_{off} b_{eq} \tag{5a}$$

$$k_{on}R(L - b_{eq}) = k_{off} b_{eq} \tag{5b}$$

which gives us the familiar solutions for the reaction kinetic equation known as the Michaelis-Menten equation. [14]

$$b_{eq_{(R \ll L)}} = \frac{RL}{L + K_D} \tag{6a}$$

$$b_{eq_{(R \gg L)}} = \frac{RL}{R + K_D} \tag{6b}$$

One of the noteworthy differences between Eq (4) and Eq (6), besides their different forms, is symmetry with respect to exchange of $R$ and $L$. The form of the solution (Eq (4)) is symmetric with respect to $R$ and $L$, which is expected given that the receptors and ligands are arbitrary labels. Therefore, interchanging the labels should not affect the final expression. On the other hand, for the special cases (Eq (6)) the asymmetry comes in because we ignore the depletion of one of the species, thereby distinguishing the labels. A symmetric expression similar to the form of equation (6) can be obtained by imposing the condition $4RL \ll (R + L + K_D)^2$ in equation 4 as

$$b_{eq_{(sym)}} = \frac{RL}{R + L + K_D}, \tag{7}$$

which reduces to equation (6) for $R \gg L$ or $R \ll L$.

Figure 1 shows the effects of the approximations in Eq (6) and Eq (7) in different regimes. We define relative error $\delta$ as

$$\delta = \frac{b_{eq} - b_{eq_{(approx)}}}{b_{eq}}$$

where $b_{eq}$ is the general form of the number of bound receptors described in Eq (4) and $b_{eq_{(approx)}}$ represents the three approximated forms for the number of bound receptors, namely $b_{eq_{(R \ll L)}}$, $b_{eq_{(R \gg L)}}$ and $b_{eq_{(sym)}}$. The relative error is plotted with respect to the number of receptor molecules $R$ while the ligand concentration $L$ and the dissociation constant $K_D$ are kept constant. Therefore, changing $R$ from $R \ll L$ to $R \gg L$ covers all the possible regimes. As expected, while Eq 6(a) and Eq 6(b) are very good approximations for Eq (4) for regimes $R \ll L$ and $R \gg L$ respectively, the relative errors for both the approximations are quite significant for other regimes. Interestingly, the symmetric expression $b_{eq_{(sym)}}$ is the most accurate approximation across all regimes.

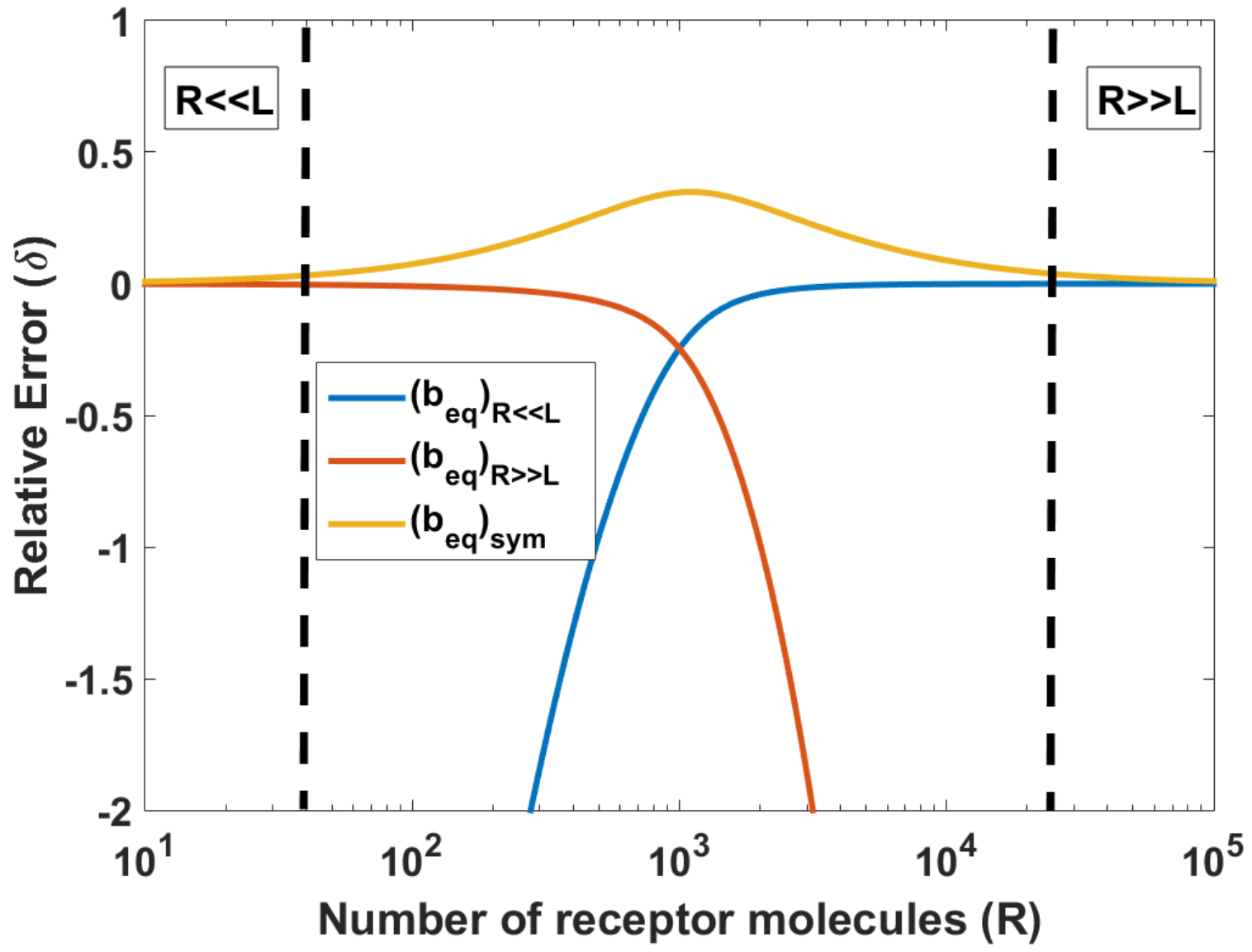

**Figure 1.** Effect of approximations: The relative error for the three approximations namely, $\left(b_{eq}\right)_{R\ll L}$, $\left(b_{eq}\right)_{R\gg L}$ and $\left(b_{eq}\right)_{sym}$ are plotted with respect to the number of receptors for all the regimes. The symmetric approximation $\left(\left(b_{eq}\right)_{sym}\right)$ shows the least error across all regimes. The relative error plots of $\left(b_{eq}\right)_{R\ll L}$ and $\left(b_{eq}\right)_{R\gg L}$ do not diverge symmetrically because the relative error for of $\left(b_{eq}\right)_{R\ll L}$ becomes independent of R for smaller values. [Parameters: $L = 1000; K_D = 100$]

### 3. Partition function for a system containing $R$ receptors and $L$ ligands

The number of bound complexes at equilibrium can also be calculated from equilibrium statistical mechanics using Gibbs-Boltzmann distribution. As before, we consider a system with $L$ ligands uniformly distributed in the solution and $R$ receptors attached to a surface. We conceptualize the $R$ receptor $L$ ligand problem in a similar manner as the one receptor $L$ ligand problem solved previously [10]. We discretize the system containing the receptor and ligand molecules in solution as being composed of $\Omega$ boxes or sites as shown in figure 2. We imagine that the discretization of the system volume into these boxes is such that any given box or site can be occupied by at most one free ligand or receptor molecule or a single ligand-receptor complex. This means that out of a total of $\Omega$ boxes, $R$ of them would constitute the ones which contain the $R$ receptor molecules present in the system, These are pictorially indicated by the blue colored boxes in figure 2. The ligand molecules, represented by the red discs in figure 2, can be distributed into any of the total $\Omega$ boxes. Note that a ligand molecule may occupy a colorless (empty) box or blue colored box. These two options represent the two states that a ligand molecule may be found in, namely, as a free molecule (if occupying a colorless box); and as a ligand-receptor complex (if occupying a blue colored box). These two states, namely, the free and bound states of the ligand molecule, are associated with energies $\epsilon_f$ and $\epsilon_b$ respectively, with the sub-scripts $f$ and $b$ denoting the free and bound states. The system may be found in different configurations each with a different total energy. For example, figure 2 shows a specific configuration for a system with $R = 7$ and $L = 12$ and number of complexes $r = 3$. The total energy $E$ of the system with the specific configuration shown in figure 2 would be $E = 9\epsilon_f + 3\epsilon_b$, because out of the 12 ligands, 9 are in the free state (occupying colorless boxes) and 3 are in the bound state (occupying blue boxes where a receptor is already present). However, as we can clearly see, there can be many more configurations which have the same value of total energy $E$ because we could rearrange the red discs into any of the other boxes which are not occupied in figure 2. Statistical mechanics provides us with a recipe to calculate the probability of finding the overall state of the system, for instance with $r = 3$ as shown in figure 2, from the number of configurations, referred to as microstates, leading to the same overall system state or total energy.

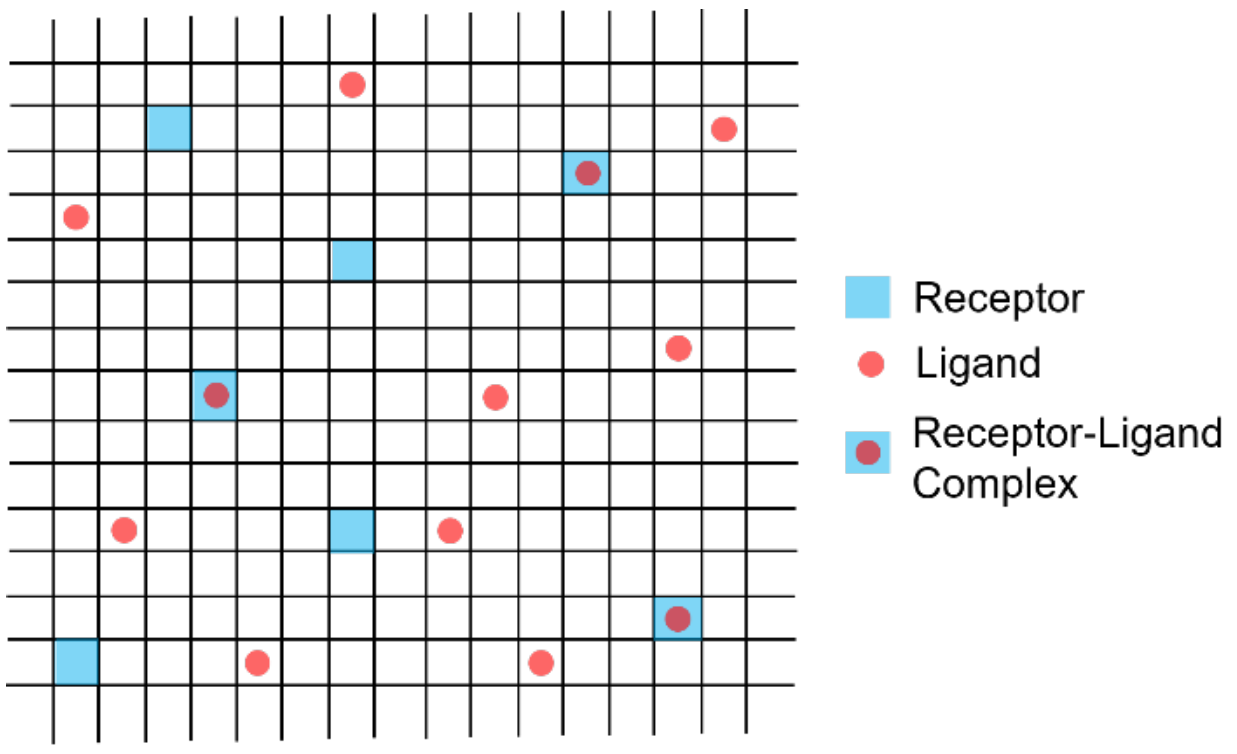

Figure 2, A schematic representation of the microstates corresponding to a state of the system. The system here is composed of 7 receptors and 12 ligands. A particular configuration corresponding to the state of the system with 3 ligand-receptor complexes is shown here. Other possible combinations are rearrangements of the objects as described in the main text which are enumerated to obtain the partition function.

Let us calculate the number of microstates for the state of the system, similar to the one shown in figure 2, where the number of receptors, ligands and complexes are $R$, $L$ and $r$ respectively. We can do this by calculating the number of ways $L$ indistinguishable ligands can be distributed among the $\Omega - R$ free sites (colorless boxes in figure 2) and the $R$ binding sites (blue boxes). The total energy of this state would be $E = (L-r)\epsilon_f + r\epsilon_b$ as there are $L - r$ ligands which are free and $r$ ligands which are bound, with energy levels $\epsilon_f$ and $\epsilon_b$ respectively. As already explained in solution to the one receptor problem, $\Omega \gg R, L$. Therefore, $\Omega - R$ can be well approximated by $\Omega$ itself. Thus, the number of arrangements or configurations of the state of the system with total energy $E = (L-r)\epsilon_f + r\epsilon_b$ is simply the number of ways to distribute $L - r$ ligands into $\Omega - R \approx \Omega$ free sites and $r$ ligands into the $R$ sites occupied by the receptors, which is simply $\binom{\Omega}{L-r}\binom{R}{r}$ where $\binom{n}{k} = \frac{n!}{k!(n-k)!}$. The partition function $Z$ is then obtained by multiplying this number of microstates with the so-called Boltzmann weight $e^{-\beta E}$, where $\beta = \frac{1}{k_B T}$ with $k_B$ the Boltzmann constant, and summing over all possible values of $r$. The smallest possible value of $r$ is clearly zero (no ligand-receptor complex) and the maximum possible value of $r$ will be the smaller of $R$ and $L$. So the partition function becomes

$$Z = \sum_{r=0}^{\min(L,R)} \binom{\Omega}{L-r} e^{-(L-r)\beta\epsilon_f} \binom{R}{r} e^{-r\beta\epsilon_b} \tag{8}$$

Equation (8) can be further simplified by invoking $\Omega \gg (L-r)$ and therefore

$$\frac{\Omega!}{(\Omega - (L-r))!} = \Omega^{L-r}. \tag{9}$$

Introducing Eq (9) into Eq (8), the expression for $Z$ can be simplified as

$$Z = \left(\Omega e^{-\beta \epsilon_f}\right)^L \sum_{r=0}^{\min(L,R)} \frac{1}{(L-r)!} \binom{R}{r} \left(\Omega e^{\beta \Delta\epsilon}\right)^{-r}, \tag{10}$$

where $\Delta\epsilon = \epsilon_b - \epsilon_f$ is the binding energy of the receptor-ligand complex.

The probability that there are $r$ bound complexes in the system can be written in terms of the partition function as,

$$p(r; L, R) = \frac{1}{Z}\left(\frac{\left(\Omega e^{-\beta \epsilon_f}\right)^L}{(L-r)!} \binom{R}{r} \left(\Omega e^{\beta \Delta\epsilon}\right)^{-r}\right) \tag{11}$$

Introducing the expression for $Z$ as derived in Eq (10) and eliminating the common terms, Eq(11) gives us

$$p(r; L, R) = \frac{\frac{R!\,L!}{(R-r)!\,(L-r)!\,r!}\left(\Omega e^{\beta \Delta\epsilon}\right)^{-r}}{\sum_{r'=0}^{\min(L,R)} \frac{R!\,L!}{(R-r')!\,(L-r')!\,r'!}\left(\Omega e^{\beta \Delta\epsilon}\right)^{-r'}} \tag{12}$$

where $r'$ is the dummy variable used as the summation index.

Equation (12) describes the general form of $p(r; L, R)$ which is easily seen to be symmetric with respect to exchange of $L$ and $R$ and hence consistent with the reaction kinetics-based results mentioned in section 1. Further, the partition function and the probability distribution shown in Eq. (11) is identical to that obtained from the master equation-based approach as expected [13]. Obtaining a closed form expression for $p(r; L, R)$ in Eq. (12) is rather difficult as illustrated in Appendix I. However, we can obtain explicit expressions for $p(r; L, R)$ for special cases where $R \gg L$ or $R \ll L$. For instance, take the case where $R \ll L$ as considered commonly in biosensor literature [15, 16, 17]. As $R \ll L$ and maximum of $r$ is the minimum of $L$ and $R$, it follows that $L \gg r$ allowing us to make the approximation,

$$\frac{L!}{(L-r)!} = L^r$$

Therefore, the partition function for this case $(Z_{R \ll L})$ becomes

$$Z_{R \ll L} = \frac{\left(\Omega e^{-\beta \epsilon_f}\right)^L (1+\alpha_L)^R}{L!}$$

where $\alpha_L = \frac{L}{\Omega e^{\beta \Delta\epsilon}}$

The probability mass function $p_{R \ll L}(r; L, R)$ can then be readily obtained as

$$p_{R \ll L}(r; L, R) = \frac{1}{(1+\alpha_L)^R}\left[\binom{R}{r} \alpha_L^r\right] \tag{13}$$

Similarly, the probability mass function for $R \gg L$ can be written by interchanging the index $L$ with $R$ in Eq (13) as

$$p_{R\gg L}(r;L,R) = \frac{1}{(1+\alpha_R)^L}\left[\binom{L}{r}\alpha_R^r\right] \tag{14}$$

where $\alpha_R = \frac{R}{\Omega e^{\beta\Delta\epsilon}}$

Unlike Eq. (12), the expressions in Eq (13) and Eq (14) are no longer symmetric with respect to the exchange of $R$ and $L$ because we are ignoring the depletion effect of one of the two species.

By re-arranging the probability mass functions described in Eq. (13) as shown in Eq. (15), we see that $p_{R\ll L}(r;L,R)$ is a Binomial distribution of the form $\binom{n}{k}p^k(1-p)^{n-k}$ with $n = R$; $k = r$ and $p = \frac{\alpha_L}{1+\alpha_L}$

$$p_{R\ll L}(r;L,R) = \binom{R}{r}\left(\frac{\alpha_L}{1+\alpha_L}\right)^r\left(\frac{1}{1+\alpha_L}\right)^{R-r} \tag{15}$$

The term $(L/\Omega)$ is the concentration of ligand molecules in the solution and the term $e^{\beta\Delta\epsilon}$ can be mapped to the macroscopic dissociation constant $K_d$ [10]. For a system of volume $v$, the dissociation constant $K_D$ described in section 2 can be written in terms of the macroscopic dissociation constant $K_d$ as $K_D = vK_d$. Introducing these relations in Eq (14), the expression for $p_{R\ll L}(r;L,R)$ can be obtained as,

$$p_{R\ll L}(r;L,R) = \binom{R}{r}\left(\frac{L}{L+K_D}\right)^r\left(\frac{K_D}{L+K_D}\right)^{R-r} \tag{16}$$

The Binomial distribution $P(k) = \binom{n}{k}p^k(1-p)^{n-k}$ for $n \to \infty$ and $p \to 0$ with the product $np$ equal to a finite value $\lambda$, can be approximated by a Poisson distribution given by $P(k) = \frac{\lambda^{-k}e^{-\lambda}}{k!}$. Here, the quantity $\lambda$ is the expectation value or the mean of the distribution denoted by $\langle k\rangle$. The number of receptor molecules in many situations is fairly large (of the order of $10^9$ or more; see section 5 for example) and we may consider $R \to \infty$. Further if $K_D \gg L$, then $p = \frac{L}{L+K_D}$ is close to zero. The product of $R$ and $p$ is $\lambda = \frac{RL}{L+K_D} \approx \frac{RL}{K_D}$. Similarly, for $R \ll L$, one can also show that the above argument is valid if $L \to \infty$ and $K_D \gg R$. Therefore, the Poisson approximation of the Binomial distribution will be given by

$$p_{K_D\gg R,L}(r;L,R) = \frac{\left(\frac{RL}{K_D}\right)^r e^{-\frac{RL}{K_D}}}{r!} \tag{17}$$

Equation 17 is commonly used in the case of multistep biological systems such as DNA transcription to simplify the analysis [18]. The expression is also relevant for describing systems where the number of bound complexes is extremely low compared to both ligand and receptors e.g. interaction of non-specific ligand molecules and cell surface receptors [12]. The steady state distribution described in Eq 17 can also be derived from the chemical master equation (CME) as described in reference [18]. To summarize, in this section we showed that under different assumptions, the number of bound complexes at equilibrium, starting from $R$ receptors and $L$ ligands follows different probability distributions. Table 1 summarizes the analysis of this section.

| Condition | Binding kinetic equation | Equilibrium Statistical Mechanics |
|---|---|---|
| $R \ll L$ | $\frac{db}{dt} = k_{on}(R-b)(L) - k_{off}b$ <br> $b_{eq} = \frac{RL}{L+K_D}$ | $p_{R\ll L}(r;L,R) = \binom{R}{r}\left(\frac{L}{L+K_D}\right)^r\left(\frac{K_D}{L+K_D}\right)^{R-r}$ <br> $\langle r\rangle = \frac{RL}{L+K_D}$ <br> $r_m = \frac{RL}{L+K_D}$ <br> **(Binomial distribution)** |
| $R \gg L$ | $\frac{db}{dt} = k_{on}(R)(L-b) - k_{off}b$ <br> $b_{eq} = \frac{RL}{R+K_D}$ | $p_{R\gg L}(r;L,R) = \binom{L}{r}\left(\frac{R}{R+K_D}\right)^r\left(\frac{K_D}{R+K_D}\right)^{L-r}$ <br> $\langle r\rangle = \frac{RL}{R+K_D}$ <br> $r_m = \frac{RL}{R+K_D}$ <br> **(Binomial distribution)** |
| $K_D \gg R, L$ | $\frac{db}{dt} = k_{on}RL - k_{off}b$ <br> $b_{eq} = \frac{RL}{K_D}$ | $p_{K_D\gg R,L}(r;L,R) = \frac{\left(\frac{LR}{K_D}\right)^r e^{-\frac{LR}{K_D}}}{r!}$ <br> $\langle r\rangle = \frac{RL}{K_D}$ <br> $r_m = \frac{RL}{K_D}$ <br> **(Poisson distribution)** |
| $R \approx L$ | $\frac{db}{dt} = k_{on}(R-b)(L-b) - k_{off}b$ <br> $b_{eq} = \frac{(R+L+K_D) - \sqrt{(R+L+K_D)^2 - 4RL}}{2}$ | $p(r;L,R) = \frac{\frac{R!\,L!}{(R-r)!\,(L-r)!\,r!}\left(\Omega e^{\beta\Delta\epsilon}\right)^{-r}}{\sum_{r'=0}^{\min(L,R)}\frac{R!\,L!}{(R-r')!\,(L-r')!\,r'!}\left(\Omega e^{\beta\Delta\epsilon}\right)^{-r'}}$ <br> $\langle r\rangle = L + \frac{K_D L\, U\,(1-L, -L+R+2, -K_D)}{U\,(-L, -L+R+1, -K_D)}$ <br> $r_m = \frac{(R+L+K_D) - \sqrt{(R+L+K_D)^2 - 4RL}}{2}$ |

Table 1: Landscape of receptor-ligand interactions: We have compiled the results obtained from the binding kinetic equation and equilibrium statistical mechanics for all the possible conditions. For the statistical mechanics method, we have the equilibrium probability distribution along with the mean ($\langle r\rangle$) and the mode ($r_m$) of the distribution which is derived in the article. One interesting point to observe here is that for all the conditions, the expression for the equilibrium solution for the binding kinetic equation $\left(b_{eq}\right)$ is equivalent to the mode $(r_m)$ of the distribution.

**4. Moments of $\boldsymbol{p(r; L, R)}$:**

How are the probability distributions of the number of complexes summarized in Table 1 related to the number of complexes in Eq. 3 and Eq. 6, derived from the reaction rate approach? To answer this question, we need to calculate the moments of the distributions. Moments are single numbers which characterize an aspect of the distribution, for instance the first moment (see equation below) of $p(r; L, R)$ gives us the expectation value or the mean number of bound receptors from the probability mass function. Let us calculate the expectation value for the case $R \ll L$, denoted by the symbol $\langle r \rangle_{R\ll L}$ from the probability mass function provided in Table 1. We can write,

$$\langle r \rangle_{R\ll L} = \sum_{r=0}^{R} r\, p_{R\ll L}(r; L, R) = \sum_{r=0}^{R} r \binom{R}{r} \left(\frac{L}{L+K_D}\right)^r \left(\frac{K_D}{L+K_D}\right)^{R-r}$$

This sum can be readily evaluated or one can use the standard result that the mean of the Binomial distribution $P(k) = \binom{n}{k} p^k (1-p)^{n-k}$ is $np$ to yield,

$$\langle r \rangle_{R\ll L} = \frac{RL}{L+K_D}, \tag{18}$$

We see that Eq. (18) is identical to Eq. 6(a) obtained from reaction rate approach. Similarly, one can also calculate $\langle r \rangle_{R\gg L}$ by interchanging $L$ and $R$ and see that the expectation value of $r$ from the probability distribution is identical to the number of complexes at equilibrium obtained from the reaction rate.

What about the Poisson approximation mentioned in Eq. (17)? We will only discuss the case $R \ll L$ as the solution to the $R \gg L$ case can be obtained by simply exchanging $R$ and $L$. As discussed in the previous section, Poisson approximation is valid when $K_D \gg L$. This means that the off-rate is significantly high compared to the depletion effect of the ligands, i.e. the ligand-receptor complexes break down into free ligands and free receptors rapidly and therefore, one may assume that $R$ receptors and $L$ ligands are always available for association. The rate of formation of complexes ignoring the depletion of receptors and ligands reads

$$\frac{db}{dt} = k_{on} RL - k_{off} b \tag{19}$$

From which, equating the LHS to zero at equilibrium gives the equilibrium number of complexes as

$$b_{eq} = \frac{RL}{K_D} \tag{20}$$

Eq. 20 is identical to the mean of the Poisson distribution. So for all these cases we see that the mean of the probability distribution of the number of ligand-receptor complexes is identical to the number of ligand-receptor complexes derived from the reaction rate approach. Let us now calculate the mean of the distribution for the general case described in equation 12.

For the general case, the expression for the mean number of bound complexes can be written as

$$\langle r \rangle = \sum_{r=0}^{\min(L,R)} r\, p(r; L, R) = \frac{\sum_{r=0}^{\min(L,R)} \frac{r\, R!\, L!}{(R-r)!\,(L-r)!\,r!} \left(\Omega e^{\beta \Delta \epsilon}\right)^{-r}}{\sum_{r=0}^{\min(L,R)} \frac{R!\, L!}{(R-r)!\,(L-r)!\,r!} \left(\Omega e^{\beta \Delta \epsilon}\right)^{-r}} \tag{21}$$

From the calculations described in Appendix I, $\langle r \rangle$ can be written as

$$\langle r \rangle = L + \frac{K_D L\, U\,(1-L, -L+R+2, -K_D)}{U\,(-L, -L+R+1, -K_D)} \tag{22}$$

where $U(a, b, z)$ is the confluent hypergeometric function of the second kind. Interestingly, the form of Eq. (22) is quite different from the result obtained using the rate equation in Eq (3). In Fig. 3a we have plotted the expression for $\langle r \rangle$ from Eq. (22) and the number of equilibrium ligand-receptor complexes from the reaction rate approach from Eq. 3 to numerically compare them. It is clear that while these expressions are very different algebraically, they match each other quite well numerically. The two expressions diverge marginally for $L \approx R$.

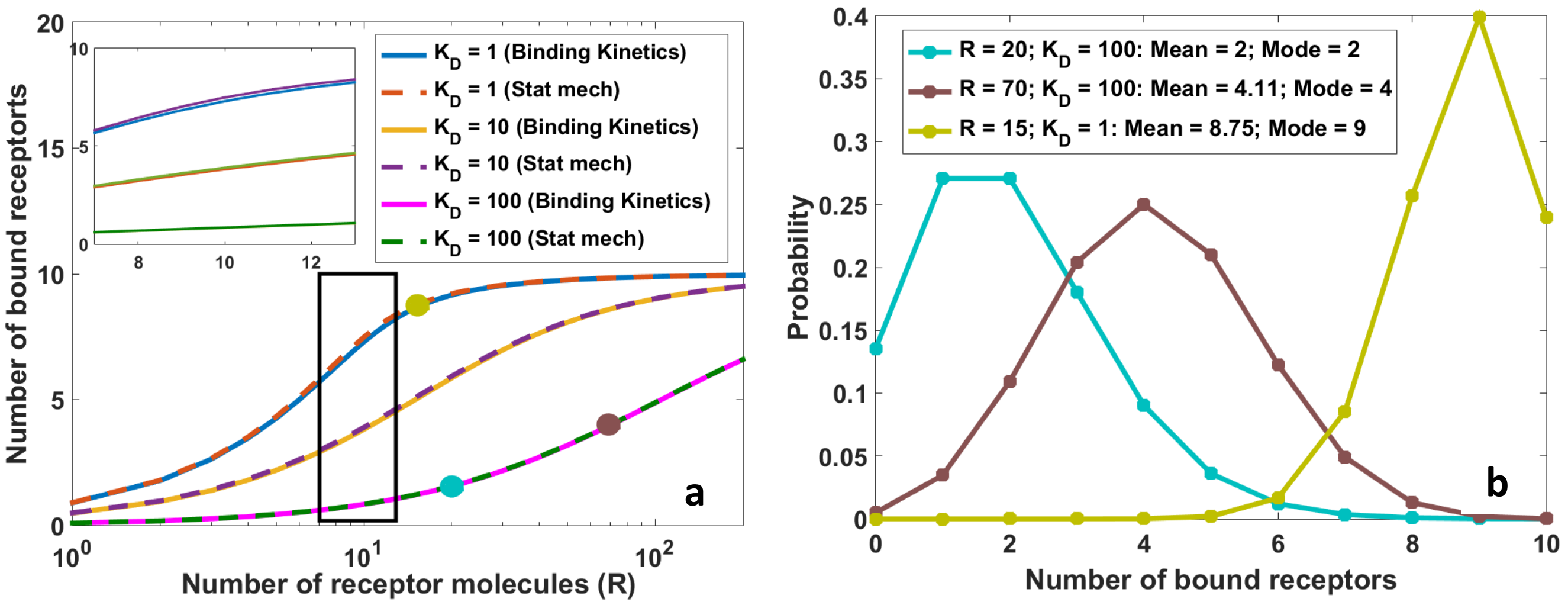

**Figure 3: a.** Comparison of the general forms of the solution obtained from the binding kinetic equation and equilibrium statistical mechanics. For special cases i.e. $|L - R| \gg 0$, both the solutions match perfectly. However, for $L \approx R$, there is some difference between them as shown in the subfigure, particularly for low values of $K_D$. The difference reduces with increase in $K_D$. **b.** Probability distribution for different conditions which are shown by coloured circles in Fig a. Irrespective of the shape of the distribution, the mean and mode are very close to each other. [Parameters: $L = 10$]

While the number of ligand-receptor complexes calculated using reaction kinetics approach numerically matches the expectation value obtained from the probability mass function of the number of complexes, the algebraic expressions are very different. Is there any other quantity derivable from the probability mass function of the ligand-receptor complexes which exactly matches the solution form the reaction rate approach? Indeed, we see that the mode of $p(r; L, R)$, denoted by $r_m$, exactly matches Eq. (3) derived from the reaction rate approach. To calculate the mode of a distribution, we need to maximize the distribution with respect to $r$. We maximize $\log\big(p(r; L, R)\big)$ taking advantage of the monotonicity of the log function.

Introducing log in equation (11), we get

$$\log p(r; L, R) = \log\left(\frac{R!\,L!}{Z}\right) - \log\big((R-r)!\big) - \log\big((L-r)!\big) - \log(r!) - r\log(\Omega e^{\beta\Delta\epsilon}) \tag{23}$$

Using Stirling's approximation

$$\log x! = x\log x - x$$

in equation (23) and maximizing with respect to $r$ gives us

$$\log\left(\frac{(L-r_m)(R-r_m)}{r_m\Omega e^{\beta\Delta\epsilon}}\right) = 0 \tag{24}$$

where $r_m$ is the mode of the probability distribution. As explained before, $\Omega e^{\beta\Delta\epsilon}$ can be mapped to the dissociation constant $K_D$ described in section 2. Therefore, Eq (24) becomes

$$(L-r_m)(R-r_m) = r_m K_D, \tag{25}$$

which can be solved for $r_m$ to obtain

$$r_m = \frac{(R+L+K_D) \pm \sqrt{(R+L+K_D)^2 - 4RL}}{2} \tag{26}$$

Similar to $r_{eq}$, the upper bound of $r_m$ is $\max(L, R)$. Therefore, we only consider the negative quadratic to be the physically acceptable solution. Therefore, $r_m$ becomes

$$r_m = \frac{(R+L+K_D) - \sqrt{(R+L+K_D)^2 - 4RL}}{2} \tag{27}$$

Therefore, the number of bound receptors at steady state $b_{eq}$ derived in section 1 is the number of bound receptors one is expected to obtain experimentally with maximum probability and not the average number of bound receptors which is commonly used in the literature [18]. Then, why does the number of bound complexes obtained from reaction rate equations match the expected value $\langle r\rangle$ obtained from the probability distribution function $p(r; L, R)$ numerically even though their analytical expressions are quite different? To answer this, we note that all the distributions in Table 1, including special cases such as $R \ll L$ have their mode close to the mean irrespective of asymmetries in the distributions. Figure 3b illustrates the point mentioned above.

**5. Higher moments of $\boldsymbol{p(r; L, R)}$:**

One advantage of the equilibrium statistical mechanics method over the reaction rate approach is that it allows us to calculate all the moments of the probability distribution function at equilibrium. Having established that $p_{R\ll L}(r; L, R)$ is the binomial distribution for the special case of $R \ll L$ considered here, we can readily find the variance in the number of bound complexes, $\sigma^2_{R\ll L}$, as

$$\sigma^2_{R\ll L} = \frac{RLK_D}{(L+K_D)^2} \tag{28}$$

Similarly, for $R \gg L$, the variance can be obtained by interchanging $L$ and $R$ in Eq 28.

Although, obtaining closed form expressions for the higher moments for the general form of $p(r; L, R)$ is not possible, we can obtain a general expression for the higher moments.

To estimate the higher moments, we define a moment generating function

$$G(z) = \sum_{r=0}^{\min(L,R)} z^r p(r; L, R) \tag{29}$$

The $n^{th}$ moment $M(n)$ of $p(r; L, R)$ can be obtained by calculating the $n^{th}$ partial differential of $G(z)$ at $z = 1$. The $n^{th}$ partial differential of $G(z)$ at $z = 1$ gives us

$$\partial_z^n G(z)|_{z=1} = \sum_{r=0}^{\min(L,R)} \left( \prod_{i=0}^{n} (r - i)\, p(r; L, R) \right) \tag{30}$$

Expanding Eq (30), we get

$$\partial_z^n G(z)|_{z=1} = \sum_{r=0}^{\min(L,R)} p(r; L, R) \left( r^n - \sum_{a=1}^{r-1} a r^{n-1} + \sum_{b=1}^{r-1} \sum_{c>b}^{r-1} bc r^{n-2} \dots \dots \right) \tag{31}$$

In literature, the coefficient of $r^k$ in equation (31) is called Stirling's number of first kind and is denoted as $s(n, k)$ [19]. Therefore, Eq (32) can be rewritten in terms of Stirling's numbers as

$$\partial_z^n G(z)|_{z=1} = \sum_{r=0}^{\min(L,R)} p(r; L, R) \sum_{k=0}^{n} s(n, k) r^k \tag{32}$$

The $n^{th}$ moment $M(n)$ now can be calculated as

$$M(n) = \partial_z^n G(z)|_{z=1} - \sum_{i=1}^{n-1} \rho(n - i) \partial_z^i G(z)|_{z=1} \tag{33}$$

where $\rho(k) = s(n, k) - \sum_{i=1}^{k-1} \rho(i) s(n - i, k - i)$

with $\rho(0) = 1$, and $\rho(1) = s(n, 1)$

**6. A numerical example: Micro/Nano particle-based molecular sensors**

Early, pre-symptomatic, detection of cancer is an unmet demand which can potentially save millions of lives annually [20]. Detection of a class of molecules called cytokines has been pursued as a means to achieve early detection of cancer [21]. In such sensors, a sample volume $V_s$ containing the target ligands interacts with a system containing receptors and the number of bound complexes translates into a physically measurable quantity such as a change in electrical conductance or optical absorbance [22]. For a designer of such a molecular sensing system, it is important to understand the effect of design choices such as the area of the sensing region used to capture target ligand molecules from sample, the volume of the sample to be used etc. on the noise due to the stochastic nature of molecular binding. Using what we learnt about the variance of molecular binding in the previous section, we can begin to answer such questions an example of which is provided below.

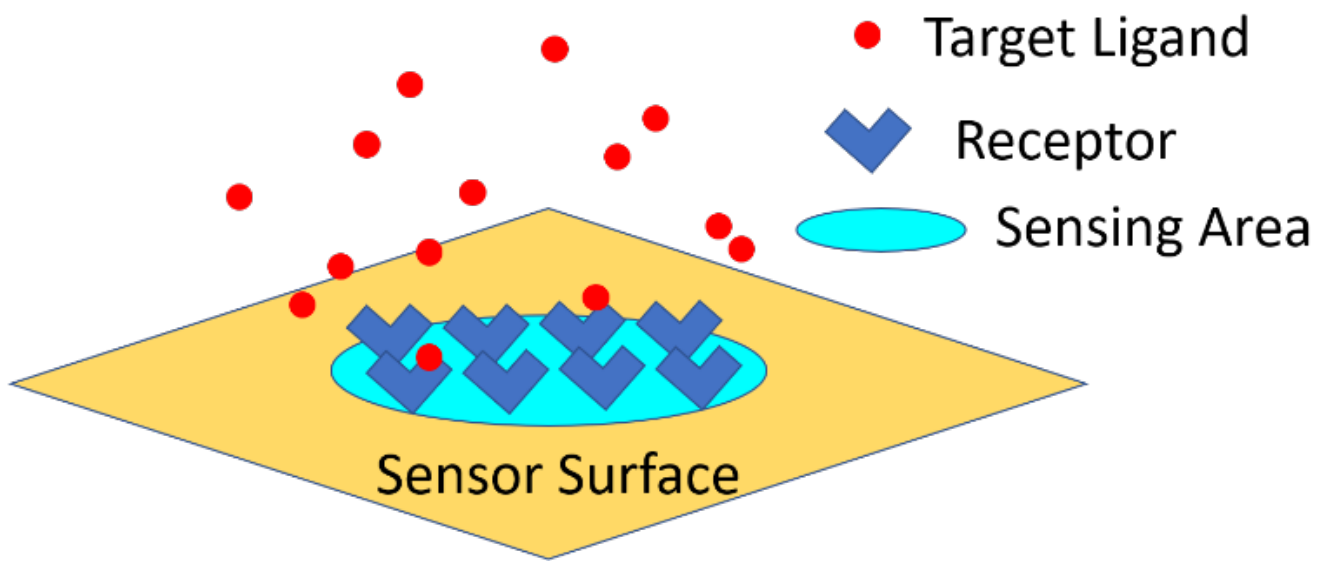

Figure 4, Schematic representation of a molecular sensing device. Receptors immobilized in a sensing area bind target ligands in a sample.

Let us consider a sensor system, as schematically shown in figure 4, consisting of a sensor surface of radius $a$ where receptor molecules are attached. Typically, the surface of the particle is saturated with a monolayer of receptor molecules. Therefore, the number of receptor molecules on each bead can be estimated to be $R = \frac{\pi a^2}{d^2}$ where $d$ is the size of a single receptor molecule which we can take to be of the order of 10 nm [23]. The number of ligand molecules is $L = cV_s$ where $c$ is the concentration of ligand molecules in the sample. Given these conditions, how should one choose the area of the sensing region (size $a$ in figure 4) so that the fluctuation in the number of the bound receptors is less than 1% of the average number of the bound receptors?

For a real system, $V_s$ is typically of the order of 100 μL. Taking $c$ to be 10 pico-molar (pM), the number of ligand molecules in the system becomes $6\text{x}10^8$. The mean number of bound receptors, μ, and the variance in the number of bound receptors, $\sigma^2$, on the sensing region can be obtained as,

$$\mu = \frac{RL}{L + K_D} \tag{34a}$$

$$\sigma^2 = \frac{RLK_D}{(L + K_D)^2} \tag{34b}$$

We define relative fluctuation per particle as $\sigma_{\text{REL}} = \frac{\sigma}{\mu}$ as

$$\sigma_{\text{REL}} = \sqrt{\frac{K_D}{RL}} \tag{35}$$

Substituting the expression for $R$ and $L$ from above, we get,

$$\sigma_{REL} = \sqrt{\frac{K_D d^2}{\pi a^2 c V_s}} \tag{36}$$

We take the value of $K_D$ to be around 6x10$^9$, corresponding to a dissociation constant of 100 pM. If the desired fluctuation of the bound number of receptors (which will generate the sensor signal) is to be less than 1%, from Eq. (37), we see that

$$a > \sqrt{\frac{K_D d^2}{\pi \sigma_{REL} c V_s}}$$

Putting the numerical values we see that the size of the sensing region must be greater than about 200 nm to achieve a relative fluctuation of less than 1% in the number of bound receptors. Why does this happen? This is because reducing the size of the sensing region decreases the number of receptors available in the sensing region and increases the relative fluctuations as seen from Eq. (36). This exercise immediately tells us that extreme miniaturization as seen in single molecule level sensors such as those based on nanopores or nano field-effect transistors [24, 25] are likely to suffer from large relative fluctuations due to the small number of receptors available.

## 7. Conclusions

In this article, we derived the equilibrium probability distribution of the number of ligand-receptor complexes in a system consisting of $R$ receptors and $L$ ligands. This work is an extension of a similar problem with one receptor and $L$ ligands which is now widely available in the references provided in this article. However, a treatment of the general $R$ receptor case and the thorough description of the connections between equilibrium probability distributions obtained and their connections with the number of complexes obtained using the reaction rate approach under various special cases is lacking in current literature. We hope that this article has addressed this gap to some extent. For instance, in Table 1, we summarized different situations relevant to molecular sensing and compared the equilibrium statistical mechanics and the reaction rate approaches and showed how these relate to one another, thus allowing the student to appreciate the equivalence of these methods. Further, this article helped us to illustrate two aspects of this problem which are not directly apparent in the solution to the single receptor problem, namely, a) a symmetry of the equilibrium probability distribution of the number of bound complexes under exchange of $R$ and $L$ and b) the number of bound complexes obtained from chemical kinetic equations has an exact correspondence to the maximum probable value of $r$. Overall, we hope that this article introduces the application of statistical mechanics to molecular binding problems in a broader setting than previous works.

## Appendix: Expression for $p(r; L, R)$ in the general case

From Abramowitz and Stegun [19] Eq 13.5.2, we obtain

$$U(a,b,z) = z^{-a}\left\{\sum_{r=0}^{T-1}\frac{(a)_r(1+a-b)_r}{r!}(-z)^{-r} + O(|z|^{-T})\right\} \qquad \text{(A1)}$$

Here $U(a,b,z)$ is the confluent hypergeometric function of second kind and $(x)_n$ is the Pochhammer symbol defined as

$$(x)_n = \frac{\Gamma(x+n)}{\Gamma(x)},$$

$\Gamma(x)$ is the Gamma function.

The summation in Eq (A1) equivalent to the partition function $Z$ defined in Eq (9) for $T = \min(L, R) + 1, a = -L,\ b = 1 + R - L,$ and $z = -K_D$. The error term can be neglected for $K_D \gg 0$ or $\min(L, R) \gg 0$. Introducing the parameters in Eq (A1) and using the property

$$(-x)_n = (-1)^n (x - n + 1)_n,$$

we obtain

$$Z = \sum_{r=0}^{\min(L,R)} \left(\frac{L!}{(L-r)!\,r!}\right)\left(\frac{R!}{(R-r)!}\right)(K_D)^{-r} = (-K_D)^{-L}\,U(-L, 1 + R - L, -K_D) \tag{A2}$$

To derive the expression for the mean of the distribution $(\mu)$, we use the moment generating function $G(\eta)$ defined in Eq (28) in the main text, which becomes

$$G(\eta) = \eta^L \frac{U\left(-L, 1 + R - L, -\frac{K_D}{\eta}\right)}{U(-L, 1 + R - L, -K_D)} \tag{A3}$$

$\mu$ can now be obtained from Eq (A3) as

$$\mu = \frac{\partial G(\eta)}{\partial \eta}\Big|_{\eta=1}$$

Using the relation [19]

$$\frac{d}{dz} U(a, b, z) = -a\, U(a + 1, b + 1, z)$$

we obtain $\mu$ as

$$\mu = L + \frac{K_D L\, U\,(1 - L, -L + R + 2, -K_D)}{U\,(-L, -L + R + 1, -K_D)} \tag{A4}$$